\documentclass[aps]{revtex4}
\usepackage{epsfig}
\usepackage{graphicx}
\usepackage{epstopdf}
\usepackage{textcomp}
\usepackage{amsmath,amssymb,color}

\parskip=\medskipamount

\newcommand{\be}{\begin{equation}}
\newcommand{\ee}{\end{equation}}
\newcommand{\eq}[1]{(\ref{#1})}
\newcommand{\fig}[1]{Fig.\ref{#1}}

\begin{document}

\title{Phase transitions in social networks inspired by the Schelling model}

\author{V. Avetisov$^{1}$, A. Gorsky$^{2,3}$, S. Maslov$^{4}$, S. Nechaev$^{5,6}$, and O. Valba$^{1,7}$}

\affiliation{$^1$N.N. Semenov Institute of Chemical Physics RAS, 119991, Moscow, Russian Federation \\ $^2$Institute of Information Transmission Problems RAS, Moscow, Russian Federation \\ $^3$Moscow Institute of Physics and Technology, Dolgoprudny 141700, Russian Federation \\ $^4$ Department of Bioengineering and Carl R. Woese Institute for Genomic Biology, University Urbana-Champaign, Urbana, Illinois, 61801 USA \\ $^5$Interdisciplinary Scientific Center Poncelet (CNRS UMI 2615), Moscow, Russian Federation \\ $^6$P.N. Lebedev Physical Institute RAS, 119991, Moscow, Russian Federation \\ $^7$Department of Applied Mathematics, National Research University Higher School of Economics, 101000, Moscow, Russian Federation}

\begin{abstract}

We propose two models of social segregation inspired by the Schelling model. Agents in our models are nodes of evolving social networks. The total number of social connections of each node remains constant in time, though may vary from one node to the other. The first model describes a "polychromatic" society, in which colors designate different social categories of agents. The parameter $\mu$ favors/disfavors connected "monochromatic triads", i.e. connected groups of three individuals \emph{within the same social category}, while the parameter $\nu$ controls the preference of interactions between two individuals \emph{from different social categories}. The polychromatic model has several distinct regimes in $(\mu,\nu)$-parameter space. In $\nu$-dominated region, the phase diagram is characterized by the plateau in the number of the inter-color connections, where the network is bipartite, while in $\mu$-dominated region, the network looks as two weakly connected unicolor clusters. At $\mu>\mu_{crit}$ and $\nu >\nu_{crit}$ two phases are separated by a critical line, while at small values of $\mu$ and $\nu$, a gradual crossover between the two phases occurs. The second "colorless" model describes a society in which the advantage/disadvantage of forming small fully connected communities (short cycles or cliques in a graph) is controlled by a parameter $\gamma$. We analyze the topological structure of a social network in this model and demonstrate that above a critical threshold, $\gamma^+>0$, the entire network splits into a set of weakly connected clusters, while below another threshold, $\gamma^-<0$, the network acquires a bipartite graph structure. Our results propose mechanisms of formation of self-organized communities in international communication between countries, as well as in crime clans and prehistoric societies.
\end{abstract}

\maketitle

\section{Introduction}

The celebrated Schelling model \cite{schelling} of spontaneous segregation in a society is one of the most popular models describing collective (social) behavior in communities of individuals. It can be formulated as follows: consider a lattice partially filled with social agents (individuals) represented by two colors, green and red. Colors designate social categories and according to the model, agents prefer to live in the surrounding of the neighbors with the same color. In the original work, Shelling considered a square lattice, where the number of neighbors is 8 (direct and diagonal nearest neighbors). An agent of some color (say, red), who has fraction of nearest neighbors of the same color below some threshold, $f$, is "unhappy" with its surrounding and can be moved to a randomly selected nearest empty lattice cell as long as such cell could be found. In another version of the model, unhappy agents could be moved to any unoccupied cell independent of the distance. Repeating many times these steps, one arrives at a dynamic equilibrium. Note that the number of agents of each color is conserved in course of such evolution. If the number of steps is tends to infinity, one can speak about the phase transition, which occurs at a critical value,  $f^*$, such that for $f<f^*$ agents form monochrome clusters with mobile boundaries. The critical threshold, $f^*=\frac{1}{2}$, is exactly known for the Schelling model on a 1D lattice. For a two-dimensional square lattice the value of $f^*$ depends on whether occupied lattice sites are above or below the percolation transition. As discussed in \cite{stat2} the result $f^*=\frac{1}{2}$ in 2D holds only below the percolation transition.

The Schelling model demonstrates how relatively weak short-ranged interactions which are encoded in the preference to live with people of the same social category, combined with mixing (ability to move to an empty place) could force global (macroscopic) changes in the system \cite{stat,stat2}. Moreover, these global changes appear abruptly as a first-order phase transition when the personal intolerance to unhappiness, quantitatively expressed in $f$, falls below some critical threshold, $f^*$.

The Schelling model in its canonical form was aimed to uncover statistical mechanisms of racial/religious/social segregation associated with resettlement of humans in communities (e.g. city neighborhoods). Thus, the model naturally depends on spatial proximity among agents. To adapt the standard Schelling model to modern times, in which distant connections mediated by social networks such as Facebook, Twitter, etc, become even more important than those in agents' local physical neighborhoods, one can consider the segregation on the topological graph as this has been done in \cite{stat3,stat4,stat5,stat6} for various modifications of Schelling-like model. Our work remaining in the same paradigm, differs from the mentioned ones (i) by detailed study of the phase transitions between various topological structures of emerging communities, which is accompanied by the spectral analysis of corresponding adjacent and Laplacian matrices of networks, and (ii) by special updating procedure which keeps fixed the degree of connections of each network node.

In our study we represent the society as a collection of agents from two social categories labeled by red and green colors, connected by some relations represented by edges in a social network. Our society is dynamic, which means that agents can create/destroy social interactions (edges). To formulate the Schelling-like model on a graph, one needs: i) to quantify the affinity for creation of social interactions between agents mediated by their social categories (node colors), and ii) to define the updating (mixing) rules. The details of our model are described in the next section, while here we provide some generic motivations behind the choice of dynamics and formulate the set of phenomena we are attempting to describe by our model.

As we demonstrate below, a variant of the Schelling model operating on a graph instead of a lattice, is capable of reproducing a rich pattern of social behaviors beyond a simple segregation (clusterization). Specifically, we impose the following modifications of the original Schelling model:
\begin{enumerate}
\item We assume that the number of social connections of each agent (the vertex degree of the social network) is strictly conserved. The degree may vary from one agent to another, however for each particular agent it is fixed and cannot be changed during the society evolution.
\item Updating rules (which replace mixing in the original Shelling model) consist of adding and removing social connections between agents (network links) under the condition of a strict conservation of a vertex degree in each network node.
\item We formulate two different models (the details are provided below) for which we introduce a concept of collective triadic interactions between individuals. In the first model we consider a "polychromatic" network whose vertices belong to the set of $M$ different colors and recoloring of nodes is prohibited. The advantage/disadvantage of specific configurations are controlled by the parameter, $\mu$, which (depending on its sign) encourages or discourages the formation of unicolor triads (i.e. connected sets of three agents from the same social category joined together), and by the parameter $\nu$, which fixes the average number of cross-color links. In the second "colorless" model, the affinity parameter, $\gamma$ is attributed to any short triadic cycle in a network (triple of vertices joined together by 3 links).
\end{enumerate}

We are interested in typical patterns formed in evolving societies (networks), reached from an entirely random initial network configuration. The social network from that perspective is the Erd\H{o}s-R\'enyi graph. The extensions of our model to scale-free networks is the subject of follow-up studies.

The paper is organized as follows. In Section II we introduce the two-parametric model of a "polychromatic" randomly evolving network with the advantage (or disadvantage) of monochrome triads and inter-color links formations. In Section III we propose a model of a "colorless" randomly evolving network favoring the formation of triangles (fully connected triads of nodes). For both models we discuss the results of our numerical simulations, provide the corresponding statistical arguments, and propose possible social interpretations of observed phenomena. In Conclusion we summarize our findings and formulate questions for future investigations.

\section{Model I: Critical behavior in polychromatic networks}

\subsection{Abstract model}

Consider a topological network, where vertices are individuals (agents), and links -- "relations between individuals". Vertices (nodes) could be of different "features", tentatively assigned as "colors". For simplicity, we consider in details a dichromatic (green-red) network of nodes, however, obtained results can be straightforwardly generalized to polychromatic networks of $M\ge 2$ colors.

The initial network configuration is a paricular random sample from a standard Erd\H{o}s-R\'enyi graph ensemble without multiple links. The vertex degree of each node is fixed at the network preparation and remains unchanged during the network evolution. We call such class of graphs the "constrained Erd\H{o}s-R\'enyi networks" (CERNs). As in the original Schelling model, we start with the case of $M=2$ colors and introduce two subclasses of nodes, denoted as "green" and "red", which could represent nations, races, religions, clans, genders, social statues, etc. of social agents. In our polychromatic model, individuals prefer to form "monochromatic triads" which are sets of three connected vertices connected by 2 links (open sets) or 3 links (closed sets). In \cite{2star} such sets are referred to as "2-star motifs". The advantage of formation of monochromatic triads of vertices replaces the original Schelling's "happiness with surroundings". To describe advantage/disadvantage of triads, we associate energies (chemical potentials) $\mu_{G}$ and $\mu_{R}$ to each unicolor green and red triad. In addition, we attribute the chemical potential $\nu$ to each cross-color pair of nodes. That enables us to define the partition function
\be
Z=\sum e^{-(\mu_G N_G + \mu_R N_R + \nu D N_{GR})}
\label{01}
\ee
where $N_G$ and $N_R$ are numbers of green and red triads of vertices? $\mu_G$ and $\mu_R$ are respective chemical potentials, $N_{GR}$ is the number of cross-color links, and $D$ is the average degree of the network. The sum in \eq{01} runs over all possible configurations in the network for fixed number of bonds under the condition that \emph{the degree of each vertex is the same as in the seed network}. Our partition function can be considered as the combination of the models suggested in \cite{crit1} (for $\nu=0$) and \cite{arenas} (for $\mu_G=0, \mu_R$=0).

\begin{figure}[ht]
\centerline{\includegraphics[width=12cm]{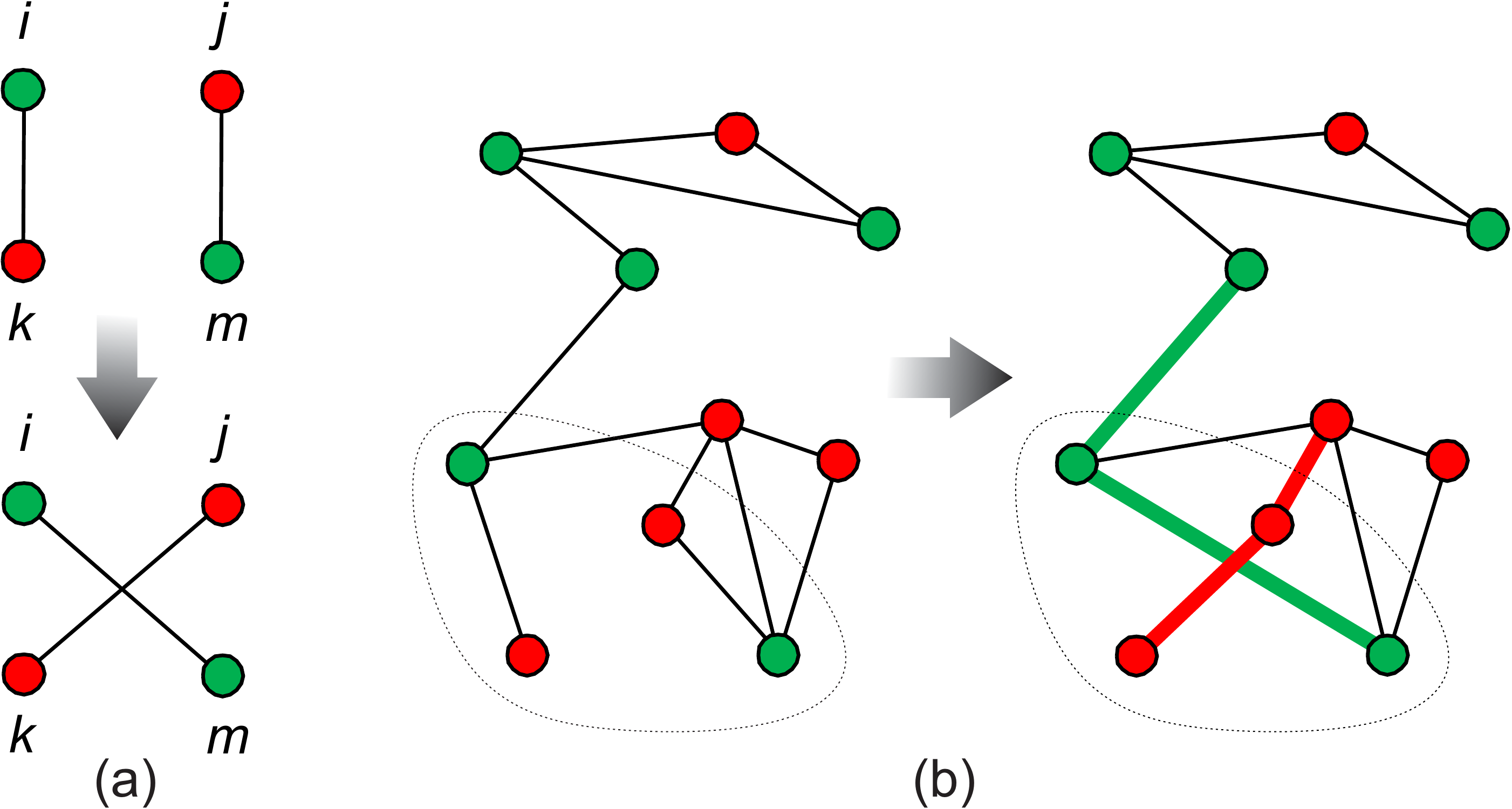}}
\caption{(a) Switching of links preserving degrees of all vertices; (b) Example of a local network updating process which increases the number of red and green triads by one. Thick lines (right) highlight edges connecting new unicolor (red and green) triads that were absent before the switch (left).}
\label{f01}
\end{figure}

Computing \eq{01} is a challenging problem, however is feasible at least in the mean-field approximation since it resembles in some aspects the generalization of the model discussed in \cite{2star}, where the interaction between nearest-neighboring \emph{nodes} is quadratic (i.e. only two joint links are involved in the interaction). The analytic approach for the two-color model will plan to discuss in a separate publication \cite{forth}. To proceed with numerics, we use currently the dynamic algorithm which replaces the numerical evaluation of the combinatorial problem \eq{01} by running stochastic evolution of the CERN (the discrete Langevin dynamics), starting from some initial configuration until the evolution converges. The initial state of the network is prepared by connecting any randomly taken pair of vertices with the probability $p$ (regardless the node color). Then, one randomly chooses two arbitrary links, say, between vertices $i$ and $j$, ($i$--$j$) and between $k$ and $m$, ($k$--$m$), and reconnect them, getting new links ($i$--$m$) and ($j$--$k$) as shown in \fig{f01}a. Such a reconnection conserves the vertex degree \cite{maslov}. As prescribed in \cite{maslov}, if at least one of the links attempted to be generated by the rewiring step exists already, this step is aborted and a new pair of links is selected. That prevents of creating multiple links between the same pair of nodes. Then, following  \cite{maslov2,maslov3}, we apply the standard Metropolis rule to each step of a reconnection. The Metropolis rules are as follows: i) if after the rewiring the number of connected unicolor connected triads of nodes (green or red) is increased, a move is accepted, ii) if the number of connected unicolor triad of nodes is decreased by some $\Delta n$, or remains unchanged, a move is accepted with the probability $e^{-\mu \Delta n}$. Here it was assumed that $\mu_R=\mu_G=\mu$, however the generalization to $\mu_R\neq \mu_G$ is straightforward. The network updating after one step of Metropolis dynamics is schematically shown in \fig{f01}b.

The Metropolis algorithm runs repetitively for a large set of randomly chosen pairs of links, until it converges. In \cite{algor} it has been shown that the algorithm actually converges to the true ground state in the equilibrium ensemble of random undirected colorless Erd\H{o}s-R\'enyi networks with fixed vertex degree. For polychromatic networks such a convergence has not yet been considered rigorously in the literature.

Below we describe typical patterns of behavior of a dichromatic constrained Erd\H{o}s-R\'enyi network in various regions of the 2D-parameter plane $(\mu,\nu)$, where $\mu=\mu_G=\mu_R$ and $\nu$ are correspondingly chemical potentials for unicolor connected triads of nodes (red and green), and for cross-color pairs of nodes.

Our investigation of a novel two-color constrained Erd\H{o}s-R\'enyi network with equal chemical potentials $\mu_G=\mu_R=\mu$ for unicolor connected triads of nodes, and $\nu$ for cross-color pairs of nodes, provides the following results in different regions of the two-dimensional parameter plane $(\mu,\nu)$.

\subsubsection{$\mu=0$: Network defragmentation in absence of affinity of unicolor connected triads}
\label{s:01}

At $\nu=0$ (compare to \cite{crit1}), the two-color network is absolutely unstable with respect to any energy $\mu>0$ favoring formation of unicolor triads of connected vertices, and immediately splits into two mostly monochrome (green and red) clusters, or "layers" (one layer = one color) with relatively small number of cross-color connections between them. The number of cross-color links, $N_{GR}$, rapidly vanishes in this regime as a function of $\mu$.

\subsubsection{$\mu\neq 0, \nu \neq 0$: Transition from "cross-community" to "intra-community" network topologies}
\label{s:02}

\begin{figure}[ht]
\centerline{\includegraphics[width=15cm]{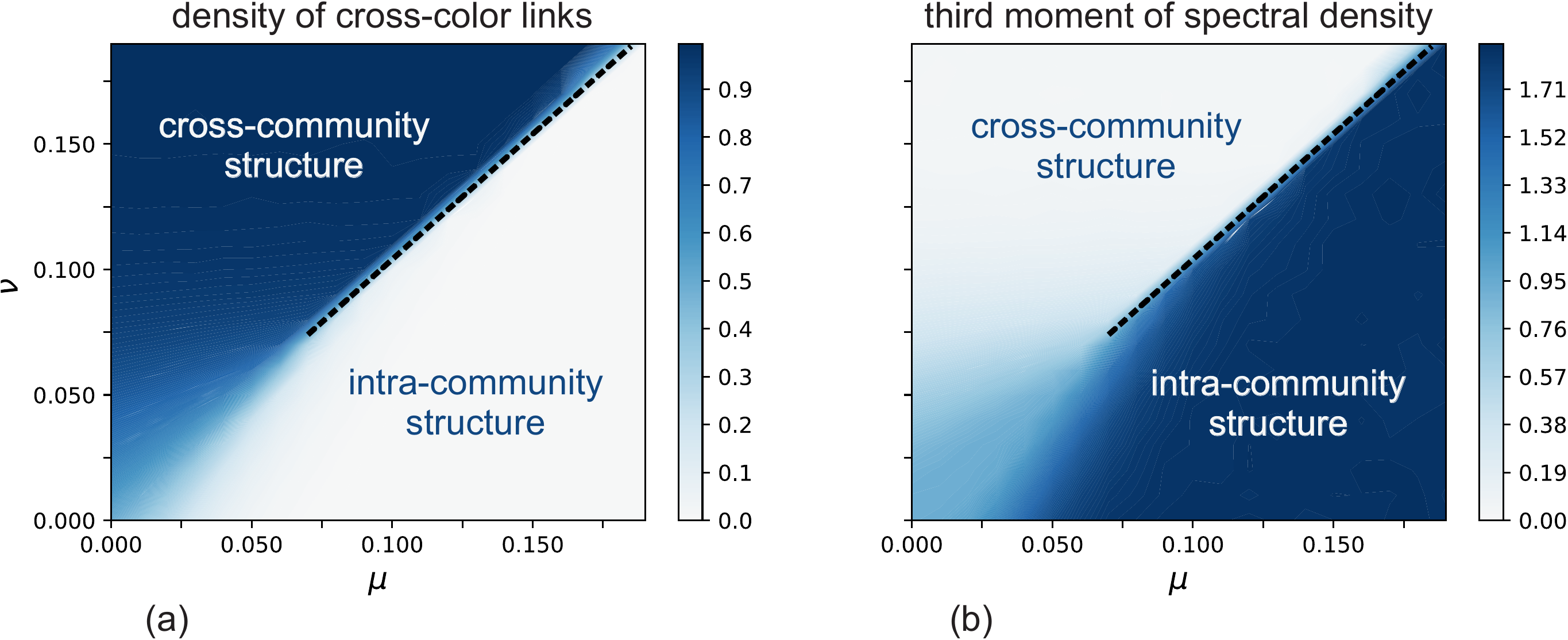}}
\caption{Phase portrait of the network in the parameter $(\mu,\nu)$--plane: (a) Density plot of the fraction of cross-color links, $\rho_{GR}(\mu,\nu)$; (b) Density plot of the third moment of the spectral density, which measures bipartitness of the network. In the simulations there are 128 nodes of each color, $p=0.15$, averaging is performed over 500 realizations.}
\label{f02}
\end{figure}

If $\mu\neq 0, \nu \neq 0$ the typical phase portrait of the model is shown in \fig{f02}a,b. In \fig{f02}a we have plotted the fraction of cross-color links, $\rho_{RG} = N_{RG}/N$. The region $\nu>\mu$ is characterized by the plateau in the density of inter-color links where the network is bipartite and demonstrates the "cross-community" structure. To the contrary, the region $\nu<\mu$ looks as two weakly connected unicolor clusters with vanishing density of cross-color links. We call this region "the intra-community dominated phase". At $(\mu>\mu_{crit},\nu>\nu_{crit})$, these two phases are separated by the first-order phase transition critical line, while at small values of $(\mu,\nu)$, the transition between two phases occurs as a smooth crossover. In the region $\nu>\mu$ at high values of $\nu$ we have $\rho_{GR}\to 1$. To understand better the topological structure of the network, we have drawn in \fig{f02}b the behavior of the third moment of the spectral density, which measures the bipartitness of the network \cite{bipart}.

\begin{figure}[ht]
\centerline{\includegraphics[width=16cm]{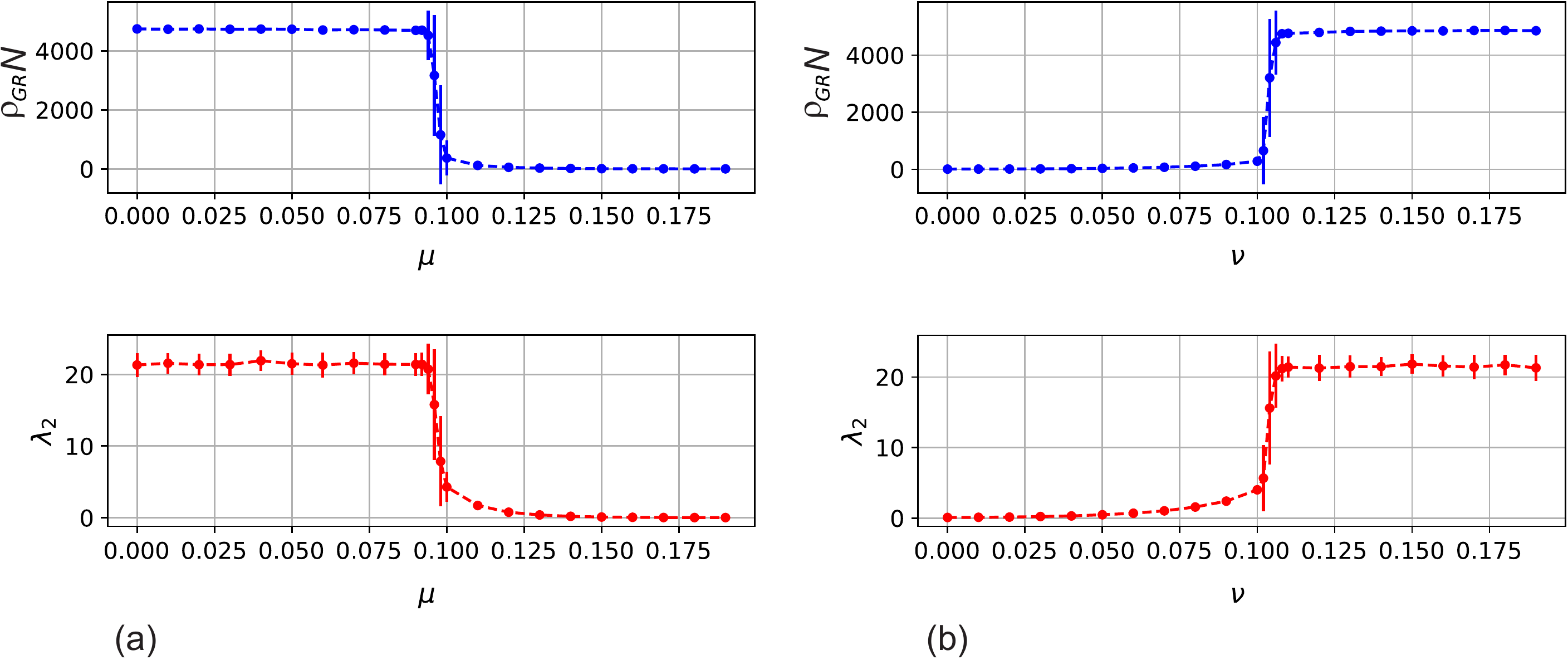}}
\caption{(a) Fraction of cross-color bonds, $\rho_{GR}$, as the function of the chemical potential of unicolor triads, $\mu$ at fixed chemical potential of cross-color bonds, $\nu=0.1$ -- upper panel, and the behavior of the second eigenvalue of Laplacian matrix, $\lambda_2(\mu)$ -- lower panel; (b) The behavior $\rho_{GR}(\nu)$ at fixed $\mu=0.1$ -- upper panel, and the behavior $\lambda_2(\nu)$ -- lower panel.}
\label{f03}
\end{figure}

The behavior of the fraction of cross-color links, $\rho_{GR}$ near the critical line is plotted in \fig{f03} for two particular cross-sections. In \fig{f03}a we have fixed the chemical potential of cross-color links, $\nu=0.1$ and have looked at the dependence $\rho_{GR}(\nu)$. In the region $\mu<\mu_{crit}$ the function $\rho_{GR}(\mu)$ develops a finite plateau, while at $\mu\approx \mu_{crit}$ the number of cross-color links nearly vanishes and remains negligible with further increasing of $\nu$. In the lower panel of \fig{f02} we have shown the dependence of the second eigenvalue, $\lambda_2$ of the Laplacian matrix of the network as a function of $\nu$. It is known that $\lambda_2$ measures the minimal number of links which should be cut to split the connected network into two disconnected parts. As one sees, the behaviors $\lambda_2(\mu)$ and $\rho_{GR}(\mu)$ coincide, which we consider as an additional support of the correct interpretation of the topological structure of our dichromatic network.

In \fig{f03}b we have studied an opposite situation and have fixed the chemical potential of unicolor triads at $\mu=0.1$, investigating the dependence $\rho_{GR}(\nu)$. Slightly below the transition point, $\nu_{crit}$, the function $\rho_{GR}(\nu)$ grows very rapidly (in fact, exponentially), reaching just above $\nu_{crit}$ the plateau, $\rho_{GR} = 1$, which is the semi-infinite plateau observed in \cite{arenas}. Measuring the dependence of the second eigenvalue $\lambda_2$ of the Laplacian matrix of the network on $\nu$, we reproduce the behavior of the function $\rho_{GR}(\nu)$ and demonstrate that in the plateau regime, the network is almost bipartite graph looking as a Corbino disk filled by the links connecting the inner and outer boundaries.

If there is no affinity of unicolor triad formation, i.e. $\mu=0$, which is the case of works \cite{arenas,radicchi,manheim}, the fraction of cross-color links $\rho_{GR}$ grows exponentially with $\nu$ slightly below $\nu_{crit}$, at which the plateau begins without the jump.

\subsubsection{Leadership formation}
\label{s:03}

\begin{figure}[ht]
\centerline{\includegraphics[width=10cm]{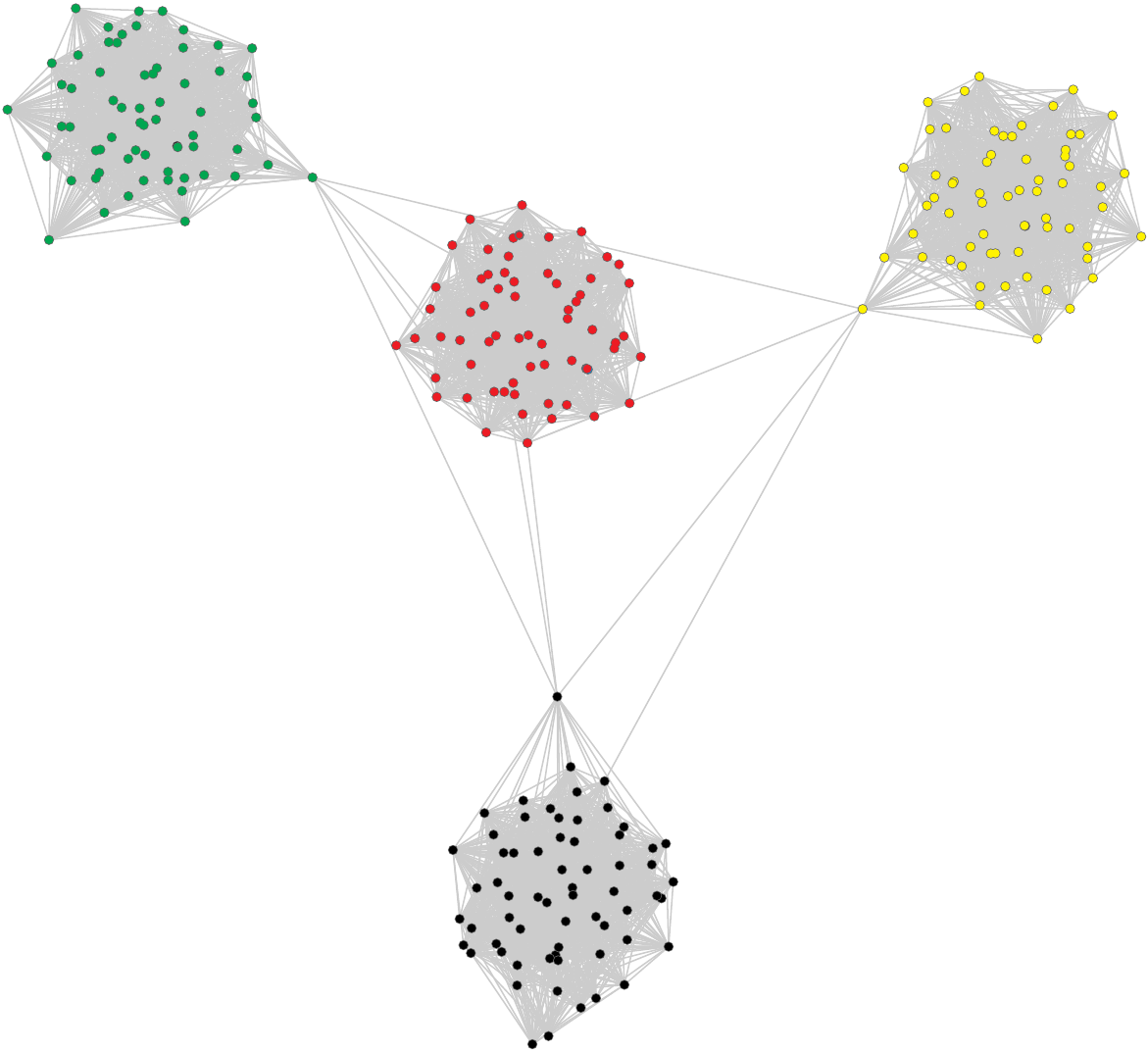}}
\caption{Typical sample of $M=4$-color network in the "intra-community" phase near the transition boundary. The "ambassadors" provide communication gates between clusters. Each cluster has 64 nodes.}
\label{f:hubs}
\end{figure}

When the number of cross-color links is small, at $\nu =\nu_{crit}$ the centrality of the network \cite{centrality} gets changed and strongly decreases above $\nu_{crit}$. The same happens at the plateau exit in the regime \ref{s:02}. In such cases the "spontaneously induced leadership" emerges. Two communities start to communicate via two spontaneously emerged "leaders" (hubs) \cite{crit2,aguirre} which have many connections inside the community and only one (or a few) emitted outside. The example of $M=4$-color network (of 64 nodes in each color) is shown  \fig{f:hubs}. The social interpretation of this effect is discussed at length of the Section \ref{s:int} while here we propose some hand-waving statistical arguments behind the phenomenon. There are two question to be elucidated:
\begin{itemize}
\item[(i)] why the clusters in the vicinity of the transition point in the "intra-community phase" ($\mu>\nu$, both $\mu$ and $\nu$ are large) communicate through "leaders" (or "ambassadors")
\item[(ii)] who becomes the "ambassador", i.e. is any correlation between the vertex degree and the possibility to become a "leader" in a polarized world?
\end{itemize}

The answer to the question (i) is straightforward. The free energy of the network consists of energetic and entropic parts. At $\mu<\mu_{crit}$ and $\nu<\nu_{crit}$ the entropy, which tends to "mix" links in the whole network as much as possible under prescribed conservation laws, has the contribution comparable to the energy. Thus, the multi-gate "democratic" communication between communities is entropically favorable. However, at $\mu>\mu_{crit}$ and $\nu>\nu_{crit}$, the entropic contribution becomes negligible with respect to the energy of collective interactions, and when $\mu<\nu$, the system tries to minimize cross-community contacts. The "ambassador"-like topology is the unique possibility for communications between polarized clusters in energy-dominated phase before the complete rupture of all relations. Whether the network could be completely deframented at $\mu\gg 1$ and finite $\nu$, depends on the density of bonds, $p$, at the preparation condition. This question definitely requires additional investigation.

The answer to the question (ii) seems a bit counterintuitive: the network nodes with intermediate values of vertex degrees (conditionally called "dark horses"), are the best candidates for the leaders in the polarized world. The statistical explanation of this effect is as follows. Network nodes, randomly acquired large vertex degree at the preparation condition are most "energetically favorable" since they could participate in many triads of the same color. From this point of view, we gain more energy connecting these nodes to the vertices of the \emph{same} color. Thus, the "best" nodes do their job inside their own communities and for communication to the "external world" the "next to the best" vertices should be chosen. It is an open and difficult question how this selection happens precisely. The sociological interpretation of this effect we provide in Section \ref{s:int}.

\subsection{Overview of the phase behavior and finite-size corrections}

The first two phenomena, described in Sections \ref{s:01}-\ref{s:02} are analogous of the segregation happen in the original Schelling model. Indeed, in the Schelling model the social segregation begins at some critical value of "happiness with the surroundings" \cite{schelling}, which in our model is replaced on the graph by the chemical potential $\mu$ controlling the affinity of connected triads of nodes of one social category. In our consideration the Schelling-like segregation occurs at any positive $\mu$.

The principal distinction between our model on a graph and the original Schelling model on a lattice is two-fold. Firstly, the adjacency matrix in the Schelling model corresponds to the distribution of neighbors on a two-dimensional surface, while in our model we deal with the topological network and do not care about lengths of network bonds and their weights, taking into account only their presence or absence. Secondly, our model on the graph demonstrates some features of spin glass behavior due to the condition of "quenched" vertex degree distribution in course of the network rearrangement, thus making the phase diagram more rich compared to the original Schelling model which by definition is "annealed".

The most important feature of the regime considered in Section \ref{s:02} is the presence of a critical line started at some point $(\mu_{crit}, \nu_{crit})$ in the phase space. This means that the passage from the "intra-community dominated phase" to the "cross-community dominated phase" at large chemical potentials occurs via a phase transition. The vertex degree conservation serves as the local constraint, while the maximization of monochromatic triads and cross-color links, is the global condition.

\begin{widetext}

\begin{figure}[ht]
\centerline{\includegraphics[width=17cm]{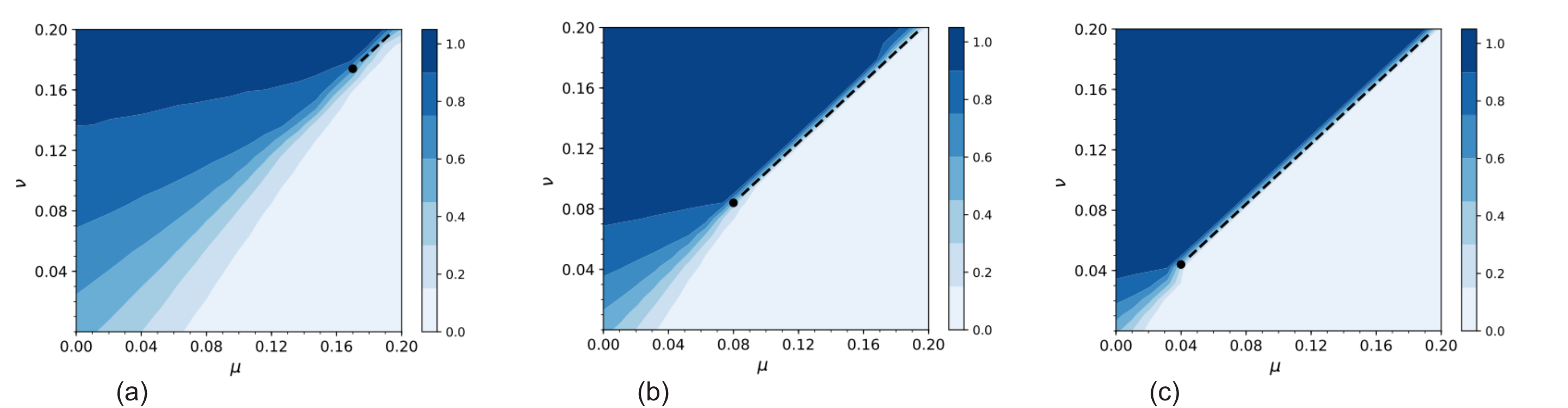}}
\caption{The phase diagrams (density plots of $\rho_{GR}$) in the $(\mu,\nu)$-plane for three different values of $N$: (a) $N=128$, (b) $N=256$, (c) $N=512$.}
\label{f04}
\end{figure}

\end{widetext}

We have investigated the dependence of the critical behavior of the system on the total number of nodes, $N$. The density plots $\rho_{GR}$ in the $(\mu,\nu)$ parametric plane are shown in \fig{f04} for three different values of $N$: for $N=128$ -- in \fig{f04}a, for $N=256$ -- in \fig{f04}b, and for $N=512$ -- in \fig{f04}c.

Two comments on these plots are in order. Firstly, we see the strong dependence of position of the end point of the critical line, $(\mu_{crit},\nu_{crit})$, on $N$: it is located at the distance $\sim N_{-1}$ along the diagonal from $(\mu,\nu)=0$. Secondly, the equation for the critical line $\mu=\nu$ is $N$-independent, though depends on the number of colors as it will be shown at length of the Section \ref{s:many}.

\subsection{Social interpretation}

In the context of social dynamics, our model captures some aspects of competition between individual and collective behaviors. This issue has been addressed in \cite{gran,competition, castellano}. The papers \cite{competition, castellano} review the applications of methods of statistical physics to the social phenomena.

The phase transition between cross-community and inter-community behavior, found in our two-parametric ($\mu,\nu$) colored topological network is expected to occur in a wide context of social systems including inter-state communications, formation of crime clans and groups of interests. We hope that some modification of the model, which takes into account different weights of directed links, could describe relationships between social agents of different genders, races, or ethnicities. The conservation of the number of connections between individuals (i.e. the conservation of degrees of  network vertices) requested in our model is crucial for the emergent behavior reported above, though seems rather natural for evolution of many real social networks. The degree of a network node represents the "extent of extroversion" (presumably fixed) of social agents incorporated into the network.

Special attention should be paid to the paper \cite{gran} discussing the situation in which agents with relatively small number of links could in some situations strongly influence the structure of the whole network. In our model the behavior discussed in \cite{gran} manifests itself in the role of "dark horses" (the bonds with moderate number of nodes): just such "weak" nodes become the leaders (or "ambassadors") in a polychromatic network in the proximity of the transition point.

It should be emphasized that our model yet describes only the non-directed connections. Thus, its immediate application to gender relations seems restricted since gender relations are not obliged to be transitive: the interest of $A$ to $B$ does not imply immediately that $B$ is interested in $A$ as well. Formally, the extension of our model to take into account the non-transitivity is not difficult -- we should attribute different weights to matrix elements $a_{ij}$ and $a_{ji}$ making the adjacency matrix non-symmetric and then apply spectral methods. However the interpretation of results is not very straightforward since the topological sense of eigenvalues (which become complex), and of moments of the spectral density for non-symmetric adjacency and Laplacian matrices is hidden. Taking into account the existence of reliable data on friendship relations (see, for example \cite{ander}), the corresponding graph-topological analysis of oriented networks and its comparison with other statistical methods is highly demanded.

\subsubsection{International relations}
\label{s:int}

The transition between the bipartite graph structure (the cross-community topology) to weakly connected closed societies (the intra-community topology) discussed in Section \ref{s:02} has possible implication in schematic description of international relations between countries, which being usually transitive, are described by the symmetric adjacency matrix. Let us begin with the "dichromatic world" consisting of agents of two colors only (red and green). At high weights of parameters $\mu$ (triple in-color connections), representing collective national relations, and $\nu$ (pair cross-color relations), representing international relations, our system can be found in one of two possible phases (i) and (ii):
\begin{itemize}
\item[(i)] at $\mu>\nu$ the network provides an example of a pair of "closed communities" characterized by a very small number of cross-color "international" links
\item[(ii)] at $\mu<\nu$ the networks acquires a nearly bipartite "open world" structure dominated by international relations.
\end{itemize}
The two phases (i) and (ii) are separated by the critical line of the phase transition with a very narrow transition region. Within this regime there is a competition between the "energies" of unicolor triads of nodes (intra-state), and pairs of cross-color nodes (inter-state). At low values of $\mu$ and $\nu$ (when $(\mu<\mu_{crit},\nu<\nu_{crit})$ -- see \fig{f02}), a crossover regime between two phases is governed by the competition between energetic and entropic effects.

One could interpret the tendency to form monochromatic triples of connected nodes, controlled by $\mu$, -- as the extent of a population nationalism, while the cross-color preference, controlled by $\nu$, -- as the tolerance towards formation of international ties. We argue that in the regime where at least one of the parameters (either $\mu$ or $\nu$) is large, a switch between the nationalism-dominated phase to the tolerance-dominated phase occurs as a 1st order phase transition. This observation might have important consequences for internal and international policies of the state. Our study clearly shows that, at high values of $\mu$ and $\nu$,  manipulating by relative weights attributed to internal and international issues is quite risky, since the switch of the population' attention between different paradigms is very sharp.

Even a modest increase of $\nu$ favoring international communications (e.g. due to state's desire to benefit from international trade), or conversely, a modest increase of $\mu$ favoring national coherence (e.g. as state's attempt to rally people around some kind of national idea, for example, national sport, or jointly fight against the "external enemy"), both could lead to a sudden and irreversible changes of the collective paradigm of the society. Societal changes at the beginning of a war may operate in this regime of an abrupt (first-order) phase transition. To the contrary, evolutionary smooth collective changes are possible at low weights in the crossover regime, when the society is less polarized.

The transition from two closed societies in a highly polarized world to the relatively open bipartite world resembles much the end of the cold war around 90s of the previous century. Our study shows that the picture of the phase transition remains the same if the chemical potentials $\mu_G$ and $\mu_R$ are different. In that case the phase transition curve depends on the sum $\mu_G+\mu_R$. In social terms it means that it is sufficient to have a strong nationalism in one country only (say, making $\mu_G$ big) to polarize the world, even in nationalism in another society is low ($\mu_R$ is small). The signature of such a behavior we see in the modern world: the patriotism based on the military rhetoric of one of current players, pushes the world to the new turn of the cold war -- formation of closed weakly communicating societies.

Even more intriguing phenomenon occurs when communities are closed and the number of inter-community links, $N_{GR}$ is relatively small. In such a regime, the importance of hubs increases drastically \cite{aguirre,crit2} culminating in the spontaneous emergence of leaders/ambassadors considered in Section \ref{s:03}. All international relations between countries pass in this case through a small number of newly created "hubs" representing country's leaders or "ambassadors".

It is eligible to ask the question: "Who becomes the ambassador at the transition point?" We have examined numerically this question and rather counterintuitive answer is as follows: the "dark horse", i.e. the "weak" node with a moderate number of links at the preparation condition becomes the leader, mediating the cross-community relations in the regime when closed societies are dominated. This rhymes well with the conclusion of the work \cite{gran} where the importance of "weak ties" has been emphasized. The statistical arguments behind this effect are provided in Section \ref{s:03}.

\subsubsection{Criminal clans}

The tendency of humans to establish social interactions with members of the same social category (the same stratum) is well known from the "everyday's experience" and is supported by many investigations. In some sense, this is the continuation of the friendship relations, which in the extreme case can be considered as collective relations in one gang. Typically friendship relations are transitive and can be attributed to the non-oriented edges connecting vertices (social agents) of the network.

The polarization of relationships between two conflicting gangs living in one area, happens according to our model, by increasing the "in-gang" collective affinity ($\mu$), decreasing of "cross-gangs" communication ($\nu$), and leads to the "intra-community" dominated network topology. In extremely polarized societies, the cross-clans relations are mediated by ambassadors -- the criminal authorities.

There are two scenarios to convert this network structure to the much less polarized bipartite "cross-community" dominated phase. The first scenario implies increasing the chemical potential of cross-community relations, $\nu$. This could be realized in practice by opening, for example, a joint sport centers where agents of different communities could meet each other. However the transition from polarized to cross-community phase is supposed to be sharp (1st order phase transition), and as every 1st order transition could be accompanied by the instability. The second scenario implies decreasing the collective in-color affinity, $\mu$, for example, by diminishing the time spent by social agents together. Such effect could be reached by providing relevant job offers. The second scenario looks more preferable since the passage to the bipartite cross-community phase happens as a crossover if $(\mu<\mu_{crit},\nu<\nu_{crit})$ and therefore is less sharp and not accompanied by instabilities.

\subsection{Spectral view on dichromatic networks}

Let us make few comments concerning the spectral properties of our dichromatic network. To this aim we need some standard notions from the graph theory. It is convenient to study the network evolution using the network adjacency matrix, $A$, of size $N\times N$, whose elements, $a_{ij}$ are defined as follows: $a_{ij}=1$ if vertices $i$ and $j$ are connected, and $a_{ij}=0$ otherwise. More information about the network structure can be obtained by studying its Laplacian matrix, $L=D-A$, which is related to the adjacency matrix, $A$, and the diagonal matrix, $D$, with the elements $d_{i}=\sum_{j=1}^{N} a_{ij}$, $i=1,...,N$. The Laplacian matrix is positively defined and has the minimal eigenvalue $\lambda_1=0$ corresponding to the homogeneous eigenvector $\mathbf{v}_1= (1,\dots, 1)$. The degeneration of $\lambda_1$ (i.e. the number of zero's eigenvalues in $L$) defines the number of disconnected components of the graph. The behavior of the second eigenvalue of the Laplacian, $\lambda_2$, in random graphs is the subject of several mathematical studies \cite{second3,second4} and has an important meaning, known as "the algebraic connectivity". In particular, if $\lambda_2>0$, the graph is connected. We have seen that $\lambda_2$ as a function of $\mu$ and $\nu$ behaves exactly the same as $N_{GR}$ see \fig{f02}, \fig{f03}. This can be considered as the check of the numerical analysis.

Quantitatively, the plateau entrance in the $\mu=0$ limit can be formulated in terms of relation between $\lambda_2$ and $\lambda_3$
\cite{arenas}.
\be
\lambda_2(A,\mu)=\lambda_3(L,\mu)
\label{05}
\ee
where $\lambda_2(A,\mu)$ is the second eigenvalue of the Laplacian matrix of the cluster $A$, and $\lambda_3(L,\mu)$ is the third eigenvalue of the Laplacian matrix of the whole network $L$. The second eigenvalue $\lambda_2$ besides its topological sense discussed in the section \ref{s:02} has an important physical meaning: it defines the diffusion time for the propagation of an excitation in the network. The relation \eq{05} is perfectly seen in our numerical simulations.

\subsection{Clusterization in polychromatic graphs}
\label{s:many}

The clusterization holds also for polychromatic networks of $M\ge 2$ colors. The unicolor clusters get formed above some critical values of $\mu_i$ which depend on $\nu_{ik}$, where $(i,j)=1,..., M$. The new phenomena taking place in the polychromatic network, is as follows. Since each cluster yields the separated eigenvalue in the spectrum of the Laplacian matrix of the network, we obtain $M$ isolated eigenvalues in the spectrum apart from the continuum. These are low-energy modes which turns out to be organized in the second zone.

\begin{figure}[ht]
\centerline{\includegraphics[width=15cm]{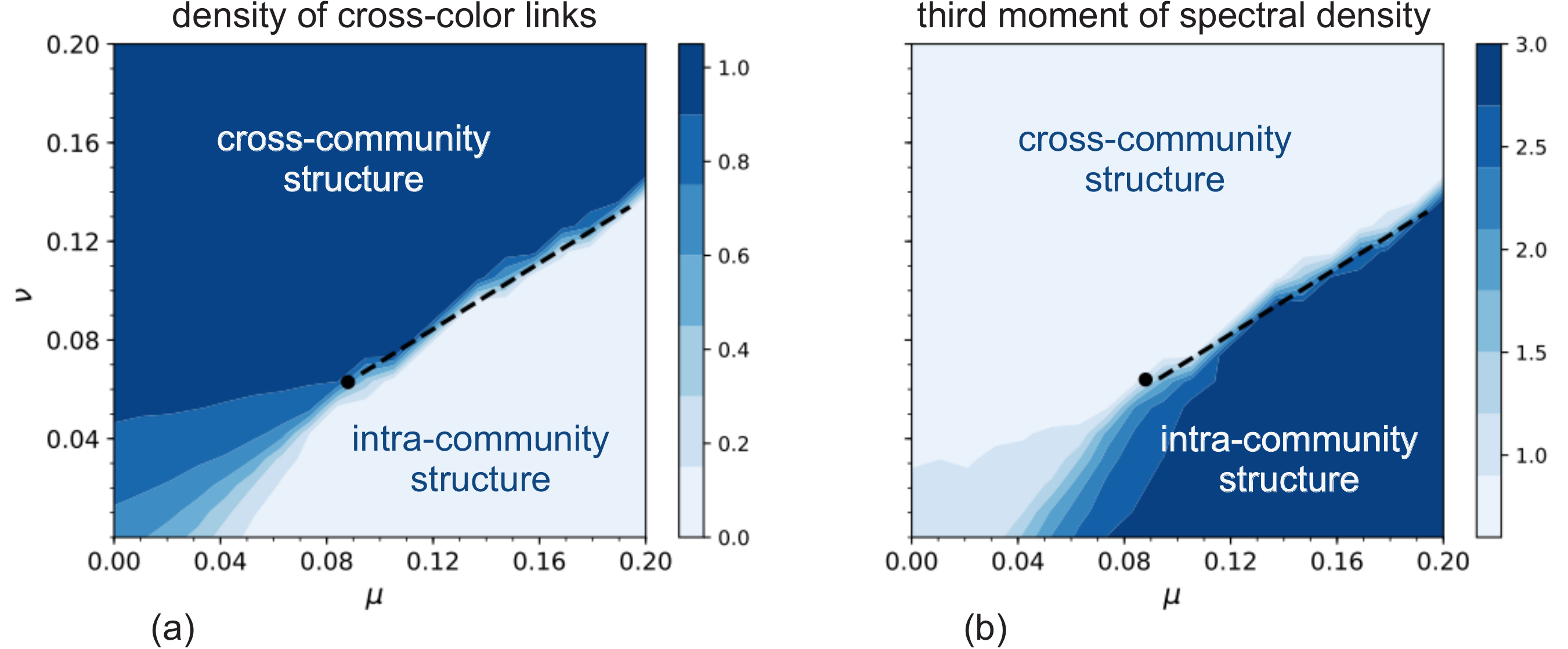}}
\caption{The density plots of phase diagram for the three-color network having 85 nodes of each color (compare to the two color network shown in \fig{f02}): (a) the cross-color density of links; (b) the averaged third moment of spectral density.}
\label{f05}
\end{figure}

To get the intuition about the critical behavior of multicolor networks we have considered the $M=3$--color network with the particular choice of chemical potentials involving only two parameters. Namely, we have simulated the network with $\mu_1=\mu_2=\mu_3=\mu$ and $\nu_{12}= \nu_{13}=\nu_{23}=\nu$. The results of simulations are presented in \fig{f05}a,b showing the density plot of cross-color links (\fig{05}a) and the third moment of the spectral density (\fig{05}b). Two phases separated by the critical lines are three monochromatic clusters and a tripartite graph. The new interesting feature of this case is that the slope of the critical line gets changed, having the slope described by the equation $\nu= \frac{2}{3} \mu$ for $M=3$. The end of the critical line approaches the point $(0,0)$ on the phase diagram in the same way as for two-color network shown in the family of diagrams \fig{f04}, i.e. as $N^{-1}$ for three networks of 85 nodes each.

The social interpretation of polychromatic network of $M$ colors is a straightforward generalization of the dichromatic model discussed at length of Section \ref{s:int}. In polychromatic networks the spontaneous emergence of communities in the initially homogeneous network is again accompanied by self-organized extraction of leaders/ambassadors. This happens when the parameter $\mu$ controlling the "co-identities of small groups" is increasing while the parameter $\nu$ of the cross-community relations is fixed. The described scenario provides a conceptual model of social dynamics in a primitive communal society when the division of labor and stratification of society just emerged.

\section{Model II: Emergence of small interconnected groups}

\subsection{Favoring of triangles (regime A)}

\begin{figure}[ht]
\centerline{\includegraphics[width=14cm]{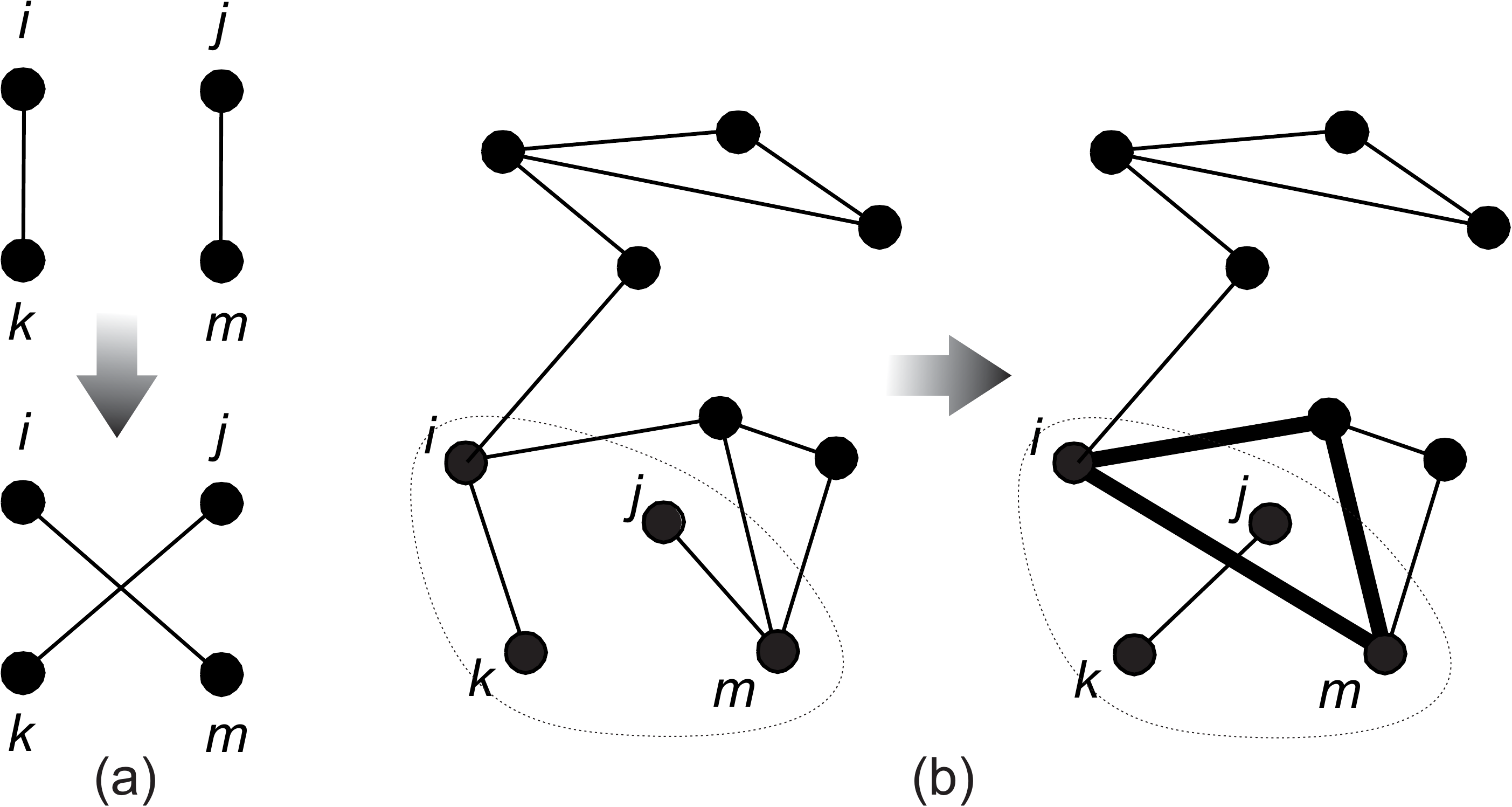}}
\caption{(a) Switching of network links preserving degrees of all vertices; (b) Example of a
local network updating move which increases the number of triangles.}
\label{f06}
\end{figure}

So far we have discussed the stability of communities in polychromatic networks with respect to the interplay between the number of unicolor triadic contacts and cross-color connections, which mimic social contacts in small groups of individuals. Here we consider a different phenomena occurring in a \emph{colorless} network, where we have an advantage/disadvantage of primitive "motifs" -- small connected subgraphs of special topology (3-cycles, 4-cycles, etc). The evolution of the network again strictly preserves the degrees of all individual nodes favoring (or preventing) the formation of triads of connected nodes. Such a degree-preserving rewiring dynamics via the Metropolis algorithm with energy determined by the number of small topological motifs (triangles, squares, or some feedback loops) was first considered in Refs. \cite{maslov2,maslov3} and is schematically depicted in \fig{f06}. The Metropolis dynamics controlled by a single chemical potential, $\gamma$, which makes formation of small closed loops ("social clubs") either preferred, or undesirable. In what follows we consider the case of triangles only (complete graph of three nodes). We are asking when the society, represented by its social network, is stable with respect to the change of $\gamma$, see \cite{avet}.

Let us begin with the case $\gamma>0$, which makes the formation of 3-cycles favorable. If the vertex degree of graph nodes is not fixed, the advantage of 3-cycles forces the formation of so-called Strauss clusters \cite{strauss}. This case has been analytically investigated in \cite{burda,newman}. It was argued that, when $\gamma$ is changing (while staying positive), the system develops two phases with essentially different concentration of 3-cycles: at large $\gamma$ the system falls into the Strauss phase with a single clique (almost full sub-graph) of nodes, while at small $\gamma$ the system looks as a weakly clustered random Erd\H{o}s-R\'enyi graph. The condensation of triads is a non-perturbative phenomenon identified in \cite{burda,newman} with the 1st order phase transition or crossover for different regions of parameter space in the framework of the mean-field cavity-like approach.

\begin{figure}[ht]
\centerline{\includegraphics[width=15cm]{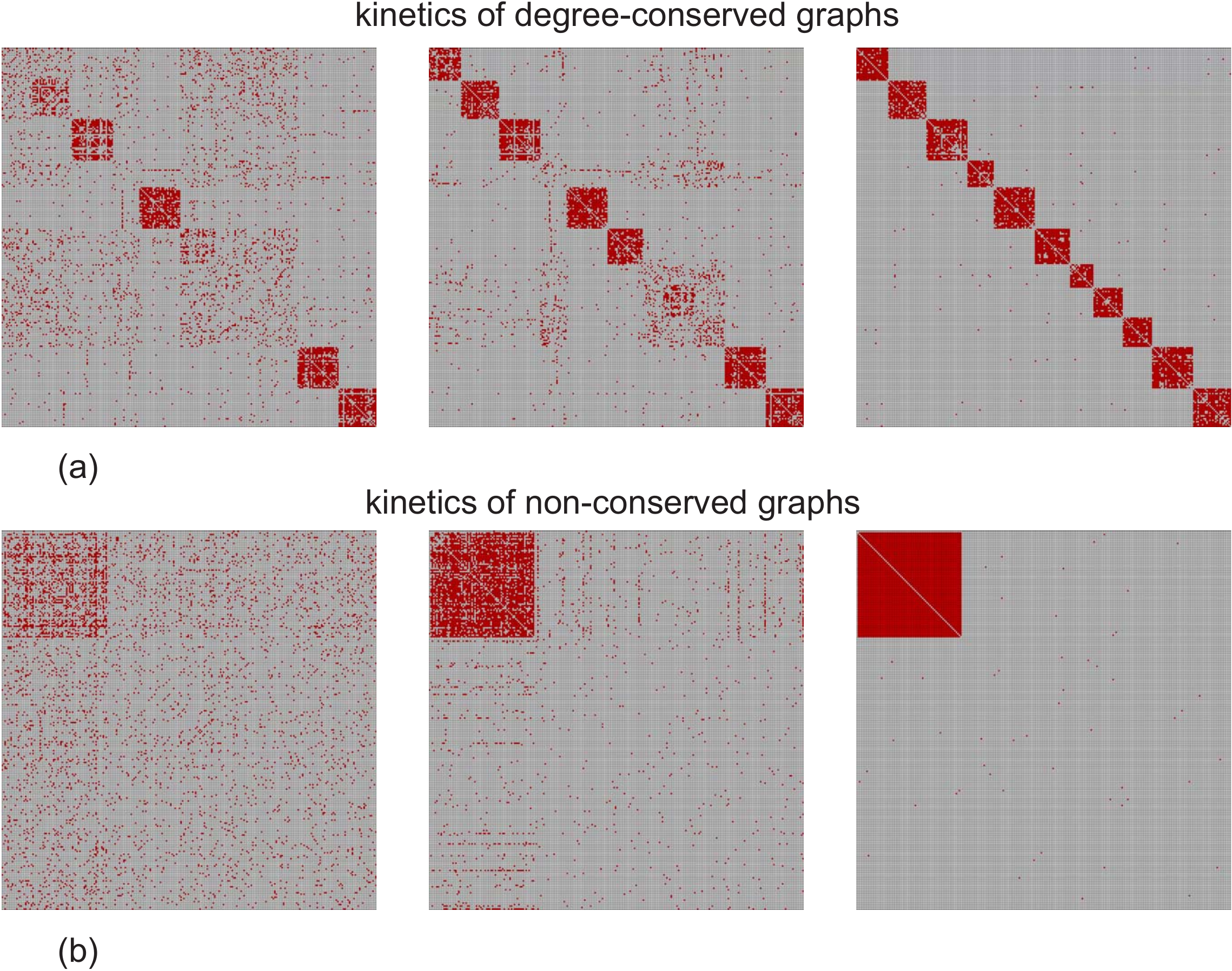}}
\caption{Few typical samples of intermediate stages of network evolution: (a) Networks
with strictly conserved vertex degree (our model); (b) Networks with non-conserved
vertex degree ("Strauss model") [reproduced from \cite{avet}].}
\label{f07}
\end{figure}

The system behaves essentially differently when a vertex degree is strictly conserved during the Metropolis rewiring. We found that above some critical fugacity, $\gamma_c$, a large network is fragmented into a collection of $[p^{-1}]$ almost fully connected sub-graphs (cliques) \cite{avet}, where $p$ is the bond formation probability in the initial random Erd\H{o}s-R\'enyi network and $[...]$ denotes the integer part of the argument. In \fig{f07}a,b we show typical structure of adjacency matrices at few intermediate stages of network rewiring towards the ground states of constrained (\fig{f07}a) and unconstrained (\fig{f07}b) Erd\H{o}s-R\'enyi networks (reproduced from \cite{avet}). The adjacency matrix $A$ in the ground state of the constrained network has block-diagonal structure with slightly fluctuating blocks of the mean size $N/[p^{-1}] \approx Np$. In contrast, the ground state in the unconstrained Erd\H{o}s-R\'enyi network in the Strauss phase consists of a single complete clique. To visualize the kinetics, we enumerated vertices at the preparation condition in arbitrary order and run the Metropolis stochastic dynamics. When the system is equilibrated and the cliques are formed, we re-enumerated vertices sequentially according to their belongings to cliques. Then we restored corresponding dynamic pathways back to the initial configuration.

It should be pointed out that the phase transition in the model with the conserved vertex degree distribution has been discussed in \cite{tamm}, however the key nature of the system, the multi-clique structure, was overlooked.

\subsection{Suppression of triangles (regime B)}

\begin{figure}[ht]
\centerline{\includegraphics[width=13cm]{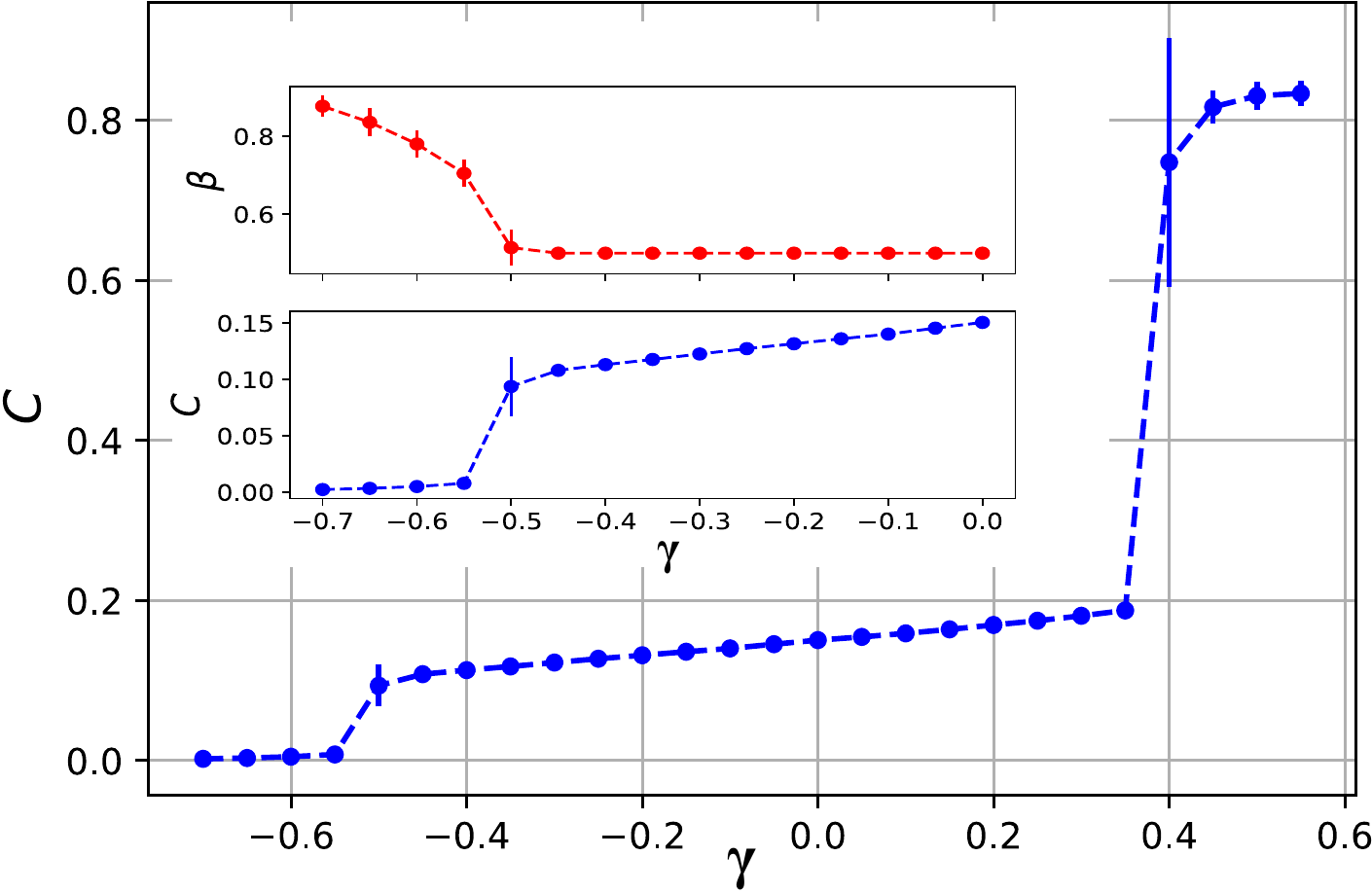}}
\caption{Critical behavior at negative $\gamma$.}
\label{f08}
\end{figure}

The negative value of the chemical potential $\gamma$ means that the triangles are suppressed. In this case the network evolution gives rise to an interesting critical behavior conjectured in \cite{bipart}, resulting in a nearly bipartite network in which connections within each cluster are suppressed. Namely, in the vicinity of some critical value, $\gamma_c<0$, the number of triangles decreases nearly to zero, making the network bipartite. The adjacency matrix acquires the block-off-diagonal form. Topologically the network looks as an annulus filled by links connecting nodes in the boundaries. In \fig{f08} the dependence of the transitivity $C$ and the Estrada index $\beta$ which are the measures of network bipartiteness \cite{estrada} is plotted.

\subsection{Social interpretation of regimes A and B}

Various ground state network topologies found in models A and B have allow different social interpretations. The simplest example of a short cycle (or more generally, of a fully-connected small clique) could be identified with a family consisting of two parents and a child. Thus, by increasing/decreasing the weight favoring short cycles (which means the relative "proximity" of relations in the family), one can increase/decrease the role of a family in the social structure of the society.

On the basis of the model A, we conjecture that by increasing the role of families, i.e. by increasing the average number of triangles, one could trigger defragmentation of initially homogeneous "proto-civilization" into a collection of weakly-connected communities. The number of communities depends on the number of connections in the proto-civilization. One could speculate about the applicability of this model to the description of segregation in prehistoric societies. It is known that the transition from the life in the open air to the life in caves increased the role of tight group and hence increased the affinity of small circles which in turn forced the separation of the entire network (society) in communities-clans. It would be interesting to incorporate our model in the identification of social and cultural communities in the archeological context reviewd in \cite{brugh}.

To the contrary, in the model B the family-like communities are suppressed and a complete restructuring of the social network happens at a critical point. Below some critical level of suppression of triangles, $\gamma_c$, the whole society gets completely polarized into two big subgroups. This emergent property is driven exclusively by the entropic forces. Connections within each of two subgroups are very loose and the major fraction of links connect vertices between opposite groups. There are various social interpretations of inter-group links, but within each of these interpretations agents prefer to develop "external relations" than to form in-group connections. Interestingly, this effect goes beyond the mean-field theoretical description. Our current understanding of this transition lags behind that of its counterpart taking place at positive $\gamma$.

\section{Conclusions}

In this work we explore critical phenomena taking place in evolving social networks. In particular, we study two models inspired by the classical Schelling model of social segregation. Both models exhibit rich collective behavior. The network evolution in our simulations starts from an Erd\H{o}s-R\'enyi random graph, and all vertex degrees are strictly conserved in the course of rewiring dynamics.

The first model describes a "polychromatic" network with vertices of different "colors" which are attributed to different social categories. The driving mechanism for the network evolution is the competition between monochromatic triads of connected vertices of the same color, and links between pairs of vertices of different colors. The phase portrait of this two-parametric model is analyzed in our study. It turns out that if connected monochromatic triads dominate, the network spontaneously splits into weakly connected clusters (one cluster per each color). Defragmentation of a network with respect to such a "color segregation" can be viewed as an effective mechanism of revealing hidden layers (stratification) in a society. Since the whole effect depends only on the sum of chemical potentials for the monochromatic triadic interactions in each color, the chemical potential in one color is capable of inducing effective interactions in other colors. In the opposite limit, where the formation of cross-color links dominates, the network develops a bipartite structure. We have identified a critical line separating these two regimes at large values of chemical potentials $\mu$ and $\nu$, while at small values the phase is replaced by a crossover behavior.

The main implications of our results for real-life social networks could be as follows. In a two-color society, whenever the weight controlling intra-community triads, or that controlling pairwise inter-community connections is high enough, a "risky" regime favoring abrupt first-order phase transitions between the "cross-community" and the "intra-community" network topologies is realized. To avoid unpredictable societal transformations ("revolutions"), both triadic and pairwise weights should be kept well below some critical values. The window for the crossover regime at small weights decreases as $N^{-1}$ as the size of the network, $N$, is increased. Another interesting feature is as follows: in the regime with small number of inter-color links, each cluster of a particular color selects a single "ambassador" (the leader), to whom all cross-society connections are delegated.

In the second (colorless) model we observe how a constrained network acquires several qualitatively different topologies separated by abrupt phase transitions. The driving force behind these transitions is the advantage/disadvantage of formation of small closed connected groups of individuals (3-cycles of graphs nodes). Above some critical value $\gamma^+>0$, attributed to 3-cycle motifs, the society gets spontaneously defragmented into a set of weakly interacting communities. Contrary, the suppression of 3-cycles, leads below $\gamma^-<0$ to formation of a polarized two-community (bipartite) structure with loose intra-community connections. Such phenomena, though expected from the mathematical studies, have been observed in the simulations for the first time and to the best of our knowledge have not yet been discussed in any physical/social context.

Our simultaneous consideration of two different models (polychromatic and colorless) is motivated by following reasons: these models represent two faces of segregation: due to multiple interactions of species of different types (in the polychromatic model I) and due to formation of subgraphs of special topology (in the colorless model II). Both models deserve detailed analytic investigation and the model I seems more feasible since the corresponding Hamiltonian is quadratic in pairs of links.

There are several natural directions for the further studies. To name but a few of them in the social context, let us point out that it is desirable to consider more general situation when the total number of links in the network, $N$, could vary. In this case the Markov chain language would be useful. It should be emphasized that the networks we consider in this study, belong to the so-called "mixed ensemble" where the number of links (and vertices) is fixed as in microcanonical ensemble, while the number of 3-cycles is controlled by the chemical potential as in the canonical ensemble. Critical properties of this ensemble differ from the ones of the canonical ensemble in which chemical potentials control both for numbers of links and 3-cycles as in \cite{strauss}, or from the microcanonical ensemble in which both these numbers are fixed \cite{radin}. Comparison of different ensembles seems very intriguing question since the full microcanonical ensemble exhibits a number of phases which still are waiting for proper identification. We also plan to relax partially the constraint of strictly fixed vertex degree in all nodes, allowing the vertex degrees to fluctuate under some control in course of network rewiring. If the control of fluctuation is entirely lost, the network will ultimately fall into the Strauss phase, however how it happens for large systems, continuously or critically, is a challenging open question. Last but not least open question deals with the connection between the topological structure of directed networks and the spectral analysis of corresponding non-symmetric adjacency matrix.

\begin{acknowledgments}
The authors are grateful for A. Andreev, D. Grebenkov, A. Poddiakov, M. Tamm and D. Ushakov for useful discussions. The work of V.A. was supported within frameworks of the state task for ICP RAS 0082-2014-0001 (state registration AAAA-A17-117040610310-6); the work of A.G. was performed at the Institute for Information Transmission Problems within the financial support of the Russian Science Foundation (Grant No.14-50-00150); O.V. thanks to Program of Fundamental Reasearch of Higher School of Economics; S.N. acknowledges the support of the RFBR grant No. 16-02-00252. V.A and S.N. are grateful to the RFBR grant No. 18-29-03167.
\end{acknowledgments}


\begin{thebibliography}{99}

\bibitem{schelling} T. Schelling, Dynamic Models of Segregation, J. Math. Sociology, {\bf 1} 143 (1971)

\bibitem{stat2} L. Dall'Asta, C. Castellano, and M. Marsili, Statistical physics of the Schelling model of segregation, J. Stat. Mech., L07002 (2008)

\bibitem{stat} D. Vincovic and A. Kirman, A physical analogue of the Schelling model, PNAS {\bf 103} 19261 (2006)

\bibitem{stat3} E. Hatna and I. Benenson, Combining segregation and integration: Schelling model dynamics for heterogeneous population, J. Artificial Societies and Social Simulation, {\bf 18} 15 (2015)

\bibitem{stat4} A.Singh and M. Haahr, Creating an adaptive network of hubs using Schelling's model, Communications of the ACM, {\bf 49} 69 (2006)

\bibitem{stat5} G. Fagiolo,M. Valente and N. Vriend, Segregation in networks, J. Econ. Beh. \& Org {\bf 64} 316 (2007)

\bibitem{stat6} A. Henry,P. Prałat  and C. Zhang, Emergence of segregation in evolving social
networks, PNAS {\bf 108} 8605 (2011)

\bibitem{2star} J.Park amd M. Newman, Solution of a 2-star model of a network, Phys. Rev E. {\bf 70} 066146 (2004)

\bibitem{crit1} V. Avetisov, A. Gorsky, S. Nechaev, and O. Valba, Spontaneous Symmetry Breaking and Phase Coexistence in Two-Color Networks,  Phys. Rev. E, {\bf 93} 012302 (2016)

\bibitem{arenas} S. Gomez, A. Diaz-Guilera, J. Gomez-Gardenes, C. J. Perez-Vicente, Y. Moreno, and A. Arenas, Diffusion dynamics on multiplex networks, Phys. Rev. Lett., {\bf 110} 028701 (2013)

\bibitem{forth} A. Andreev, V. Avetisov, A. Gorsky, S. Maslov, S. Nechaev, and O. Valba, in preparation

\bibitem{maslov} S. Maslov and K. Sneppen, Specificity and Stability in Topology of Protein
Networks, Science, {\bf 296} 910 (2002)

\bibitem{maslov2} S. Maslov, K. Sneppen,  and A. Zaliznyak, Detection of topological patterns in complex networks: correlation profile of the Internet, Physica A, {\bf 333} 529 (2004)

\bibitem{maslov3} S. Maslov, K. Sneppen, and U. Alon, Correlation Profiles and Circuit Motifs in Complex Networks, an invited book chapter in  the "Handbook of Graphs and Networks", S. Bornholdt, and H.-G. Schuster, (eds.), (Wiley-VCH: Weinheim, 2003)

\bibitem{algor} F. Viger and M. Latapy, Efficient and simple generation of random simple connected graphs with prescribed degree sequence, Computing and Combinatorics, Lecture Notes in Computer Science, {\bf 440} (2005)

\bibitem{radicchi} F. Radicchi, Driving interconnected networks to supercriticality Phys. Rev. X, {\bf 4} 021014 (2014)

\bibitem{centrality} L. Freeman, A set of measures of centrality based on betweenness, Sociometry, {\bf 40} 35 (1977)

\bibitem{manheim} F. Darabi Sahneh, C. Scoglio and P. Van Mieghem, Exact coupling threshold for structural transition reveals diversified behaviors in interconnected networks, Phys. Rev. E, {\bf 92} 040801(R) (2015)

\bibitem{aguirre} J. Aguirre, R. Sevilla-Escoboza, R. Gutierrez, D. Papo, and J. M. Buldu, Synchronization of Interconnected Networks: The Role of Connector Nodes, Phys. Rev. Lett., {\bf 112} 248701 (2014)

\bibitem{crit2} V. Avetisov, A. Gorsky, S. Nechaev, and O. Valba, Finite plateau in spectral gap of polychromatic constrained random networks,  Phys. Rev. E {\bf 96} 062309 (2017)

\bibitem{gran} M. Granovetter, The srtrength of weak ties, Am. J. Sociology, {\bf 79} 1360 (1973)

\bibitem{ander} C. J. Anderson, S. Wasserman, and B. Crouch, A $p^*$ primer: logit models for social networks, Social Networks, {\bf 21} 37  (1999)

\bibitem{competition} S. Grauwin, E. Bertin, R. Lemoy, and P. Jensen, Competition between collective and individual dynamics, PNAS, {\bf 106} 20622 (2009)

\bibitem{castellano} C Castellano, S Fortunato, and V Loreto, Statistical physics of social dynamics, Rev. Mod. Phys., {\ bf 81} 591 (2009)

\bibitem{homo} M. McPherson, L. Smith-Lovin, and J.M. Cook, Birds of a feather: Homophily in social networks, Annual review of sociology, {\bf 27} 415 (2001)

\bibitem{homo2} S.M. Goodreau, J.A. Kitts, and M. Morris, Birds of a feather, or friend of a friend? Using exponential random graph models to investigate adolescent social networks, Demography, {\bf 46} 103 (2009)

\bibitem{second3} J. Friedman, On the second eigenvalue and random walks in random d-regular graphs, Combinatorica, {\bf 11(4)} 331 (1991).

\bibitem{second4} J. Friedman, A proof of Alon's second eigenvalue conjecture and related problems, Mem. Amer. Math. Soc., {\bf 910} (2008

\bibitem{avet} V. Avetisov, M. Hovanessian, A. Gorsky, S. Nechaev, M. Tamm, and O. Valba, Eigenvalue tunnelling and decay of quenched random networks, Phys. Rev. E, {\bf 94} 062313 (2016)

\bibitem{strauss} D.Strauss, On a General Class of Models for Interaction, SIAM Rev., {\bf 28} 513 (1986)

\bibitem{burda}  Z. Burda, J. Jurkiewicz, and A. Krzywicki, Network transitivity and matrix models, Phys. Rev. E
69, 026106

\bibitem{newman} J. Park and M. E. J. Newman, Solution for the properties of a clustered network, Phys. Rev. E, {\bf 72} 026136 (2005)

\bibitem{tamm} M.V. Tamm, A.B. Shkarin, V.A. Avetisov, O.V. Valba, and S.K. Nechaev, Islands of stability in motif distributions of random networks, Phys. Rev. Lett., {\bf 113} 095701 (2014)

\bibitem{newman2} B.Karrer  and M.Newman, "Random graphs containing
arbitrary distributions of subgraphs", 2010, Phys.Rev.E 82:066118

\bibitem{bipart} S. Chatterjee and P. Diaconis, Estimating and understanding exponential random graph models. arXiv: 1102.2650; C. Radin and M. Yin, Phase transitions in exponential random graphs, Annals of Applied Probability, {\bf 23} 2458 (2013)

\bibitem{estrada} E. Estrada, Spectral scaling and good expansion properties in complex networks Europhys. Lett., {\bf 73} 649 (2006)

\bibitem{brugh} T. Brughmans, Thinking Through Networks: A Review of Formal Network Methods in Archaeology, J Archaeol. Method Theory {\bf 20} 623 (2013)

\bibitem{radin} C. Radin, L. Sadun , J.Phys A, 305002 46 (2013); R. Kenyon, C. Radin, K. Ren, L. Sadun J. Stat. Phys. 233 168 (2017); R. Kenyon, C. Radin, K. Ren, L. Sadun J. Stat Phys. A435001 50 (2017)

\end{thebibliography}
\end{document}